\documentclass[aps,showpacs,prd,twocolumn]{revtex4}%
\usepackage{graphicx}
\usepackage{amssymb}
\usepackage{amsmath}
\usepackage{color}
\usepackage{float}
\usepackage{accents}
\usepackage{graphicx}
\usepackage{graphicx}
\usepackage{amsfonts}%
\providecommand{\U}[1]{\protect\rule{.1in}{.1in}}
\begin{document}
\title{A new CPT-even and Lorentz-Violating nonminimal coupling in the Dirac equation }
\author{R. Casana$^{a}$, M. M. Ferreira Jr$^{a}$, E. O. Silva$^{a}$, E. Passos$^{b}$,
F. E. P. dos Santos$^{a}$\thanks{e-mails: rodolfo.casana@gmail.com,
manojr07@ibest.com.br,}}
\affiliation{$^{a}${\small {Departamento de F\'{\i}sica, Universidade Federal do
Maranh\~{a}o, Campus Universit\'{a}rio do Bacanga, S\~{a}o Luiz - MA,
65085-580 - Brazil}}}
\affiliation{$^{b}${\small {Departamento de F\'{\i}sica, Universidade Federal de Campina
Grande, }Caixa Postal 10071, 58109-970 Campina Grande, PB, Brazil}}
\affiliation{}

\begin{abstract}
In this work, we propose a CPT-even and Lorentz-violating dimension-five
nonminimal coupling\ between fermionic and gauge fields, involving the
CTP-even and Lorentz-violating gauge tensor of the SME. This nonminimal
coupling modifies the Dirac equation, whose nonrelativistic regime is governed
by a Hamiltonian which induces new effects, such as an electric-Zeeman-like
spectrum splitting and an anomalous-like contribution to the electron magnetic
moment, between others. Some of these new effects allows to constrain the
magnitude of this nonminimal coupling in $1$\ part in $10^{16}.$

\end{abstract}

\pacs{11.30.Cp, 12.20.-m, 12.60.-i}
\maketitle

\section{Introduction}

The Standard Model Extension (SME) \cite{Colladay, Samuel} has been the usual
framework for investigating signals of Lorentz violation in physical systems.
The SME is the natural framework for studying properties of physical systems
with Lorentz-violation since it includes Lorentz-violating terms in all
sectors of the minimal standard model. The Lorentz-violating (LV) terms are
generated as vacuum expectation values of tensors defined in a high energy
scale. The SME is a theoretical framework which has inspired a great deal of
investigations in recent years. Such works encompass several distinct aspects
involving: fermion systems and radiative corrections \cite{fermion,fermion2},
CPT- probing experiments \cite{CPT}, the electromagnetic CPT- and
Lorentz-odd\ term \cite{Adam,Cherenkov1,photons1}, the nineteen
electromagnetic CPT-even and Lorentz-odd coefficients
\cite{KM1,Risse,Cherenkov2}. \ Recently, some studies involving higher
dimensional operators have also been reported with great interest
\cite{Kostelec,Myers,Mariz}. These many contributions have elucidated the
effects induced by Lorentz violation and served to set up stringent upper
bounds on the LV coefficients.

Some time ago, a Lorentz-violating and CPT-odd nonminimal coupling between
fermions and the gauge field was proposed \cite{NM1} at the form%
\begin{equation}
D_{\mu}=\partial_{\mu}+ieA_{\mu}+i\frac{g}{2}\epsilon_{\mu\lambda\alpha\beta
}(k_{AF})^{\lambda}F^{\alpha\beta},\label{cov}%
\end{equation}
in the context of the Dirac equation, $(i\gamma^{\mu}D_{\mu}-m)\Psi=0.$ Here,
the fermion spinor is $\Psi$, while $(k_{AF})^{\mu}=($v$_{0},\mathbf{v)}$ is
the Carroll-Field-Jackiw four-vector, and $g$ is the constant that measures
the nonminimal coupling magnitude. The analysis of the nonrelativistic limit
revealed that this coupling provides a magnetic moment $\left(  g\mathbf{v}%
\right)  $ for uncharged particles \cite{NM1}, yielding an Aharonov-Casher
phase for its wavefunction. It was also shown that this particular nonminimal
coupling induces topological phases in more general contexts \cite{NM3}. Its
effects on the Hydrogen spectrum were studied in Ref. \cite{NMhydrog}, while
its influence in the dynamics of the Aharonov-Bohm-Casher problem was analyzed
in Ref. \cite{NMABC}. Recently, this coupling was considered in the context of
fermion-fermion ultrarelativistic scattering, with the evaluation of
corrections to the corresponding cross section and establishment of an upper
bound as tight as $1$ part in $10^{12}$ \cite{NMmaluf}. Another recent work in
this context was developed involving aspects related to the Hall effect and
Landau levels \cite{NMhall}. Lastly, generalized versions of nonminimal
couplings have been proposed to examine the induction of several types of
topological and geometrical phases \cite{NMbakke}.

In the present work, we propose a new CPT-even,\ dimension-five, nonminimal
coupling linking the fermionic and gauge fields in the context of the Dirac
equation. By considering the nonrelativistic limit of the modified Dirac's
equation, we explicitly evaluate the new contributions to the nonrelativistic
Hamiltonian. These new terms imply a direct correction on the anomalous
magnetic moment, a kind of electrical Zeeman-like effect on the atomic
spectrum, and a Rashba-like coupling term. These effects are then used to
impose upper bounds on the magnitude of the nonminimally coupled LV
coefficients at the level of $1$ part in $10^{16}.$

\section{A CPT-even Lorentz-violating nonminimal coupling}

We consider a nonminimal coupling involving fundamental Dirac fermions and the
electromagnetic field in the context of the Dirac equation,
\begin{equation}
(i\gamma^{\mu}D_{\mu}-m_{e})\Psi^{\left(  e\right)  }=0,\label{Dirac1}%
\end{equation}
where $\Psi^{\left(  e\right)  }$ is the electron spinor wave function,
and\textbf{ }the covariant derivative with nonminimal coupling is%
\begin{equation}
D_{\mu}=\partial_{\mu}+ieA_{\mu}+\frac{\lambda^{\left(  e\right)  }}{2}\left(
K_{F}\right)  _{\mu\nu\alpha\beta}\gamma^{\nu}F^{\alpha\beta},\label{covader}%
\end{equation}
with $\left(  K_{F}\right)  _{\mu\nu\alpha\beta}$ being the tensor ruling the
Lorentz violation in the CPT-even electrodynamics of the SME, while
$\lambda^{\left(  e\right)  }$ is the electron nonminimal coupling constant.
This tensor possess 19 components, whose properties and effects have been
examined since 2002 \cite{KM1}. The background tensor $\left(  K_{F}\right)
_{\alpha\nu\rho\varphi}$ has the same symmetries of the Riemann tensor,
$\left(  K_{F}\right)  _{\alpha\nu\rho\varphi}=-\left(  K_{F}\right)
_{\nu\alpha\rho\varphi},$ $\left(  K_{F}\right)  _{\alpha\nu\rho\varphi
}=-\left(  K_{F}\right)  _{\alpha\nu\varphi\rho},$ $\left(  K_{F}\right)
_{\alpha\nu\rho\varphi}=\left(  K_{F}\right)  _{\rho\varphi\alpha\nu},$and a
double null trace, $\left(  K_{F}\right)  ^{\alpha\beta}{}_{\alpha\beta}=0,$
implying 19 components. This tensor $\left(  K_{F}\right)  _{\alpha\nu
\rho\varphi}$ can be written in terms of four $3\times3$ matrices $\kappa
_{DE},\kappa_{DB},\kappa_{HE},\kappa_{HB},$ defined in Refs. \cite{KM1} as:
\begin{align}
\left(  \kappa_{DE}\right)  _{jk} &  =-2\left(  K_{F}\right)  _{0j0k}%
,\label{Par1}\\
\text{ }\left(  \kappa_{HB}\right)  _{jk} &  =\frac{1}{2}\epsilon
_{jpq}\epsilon_{klm}\left(  K_{F}\right)  _{pqlm},\\
\text{ }\left(  \kappa_{DB}\right)  _{jk} &  =-\left(  \kappa_{HE}\right)
_{kj}=\epsilon_{kpq}\left(  K_{F}\right)  _{0jpq}.\label{Par2}%
\end{align}
The symmetric matrices $\kappa_{DE},\kappa_{HB}$ contain the parity-even
components and possess together eleven independent components, while
$\kappa_{DB},\kappa_{HE}$ possess no symmetry, having together eight
components, representing the parity-odd sector of the tensor $\left(
K_{F}\right)  $.

The Dirac equation (\ref{Dirac1}) can be explicitly written as
\begin{equation}
\left[  i\gamma^{\mu}\partial_{\mu}-e\gamma^{\mu}A_{\mu}+\frac{\lambda
^{\left(  e\right)  }}{2}\left(  K_{F}\right)  _{\mu\nu\alpha\beta}\sigma
^{\mu\nu}F^{\alpha\beta}-m_{e}\right]  \Psi^{\left(  e\right)  }%
=0,\label{DiracM1}%
\end{equation}
and
\begin{equation}
\sigma^{\mu\nu}=\frac{i}{2}(\gamma^{\mu}\gamma^{\nu}-\gamma^{\nu}\gamma^{\mu
})=\frac{i}{2}[\gamma^{\mu},\gamma^{\nu}].\label{OP1}%
\end{equation}
Thus, the relevant\ electron Lagrangian is\textbf{ }%
\begin{equation}
\mathrm{{\mathcal{L}}}_{(e)}=\bar{\Psi}^{(e)}(i{\rlap{\hbox{$\mskip 1 mu /$}}}%
\partial-e{\rlap{\hbox{$\mskip 1 mu /$}}}A-m_{e}+\frac{\lambda^{\left(
e\right)  }}{2}\left(  K_{F}\right)  _{\mu\nu\alpha\beta}\sigma^{\mu\nu
}F^{\alpha\beta})\Psi^{(e)}.\label{Le}%
\end{equation}
Using the parametrization (\ref{Par1})-(\ref{Par2}), we obtain:
\begin{align}
&  \left(  K_{F}\right)  _{\mu\nu\alpha\beta}\sigma^{\mu\nu}F^{\alpha\beta
}=2\sigma^{0i}\left[  \left(  \kappa_{DE}\right)  _{ij}E^{j}+\left(
\kappa_{DB}\right)  _{ij}B^{j}\right]  \nonumber\\
&  \hspace{1.5cm}+\epsilon_{kij}\sigma^{ij}\left[  \left(  \kappa_{HE}\right)
_{kq}E^{q}+\left(  \kappa_{HB}\right)  _{kq}B^{q}\right]  ,\label{KF}%
\end{align}
where we have used $F_{0j}=E^{j},F_{mn}=\epsilon_{mnp}B^{p}$, and $\sigma
^{0i}$, $\sigma^{ij}$ are the components of the operator (\ref{OP1}), \
\begin{equation}
\sigma^{0i}=i\left(
\begin{array}
[c]{cc}%
0 & \sigma^{i}\\
\sigma^{i} & 0
\end{array}
\right)  ,\text{ \ \ }\sigma^{ij}=-\left(
\begin{array}
[c]{cc}%
\epsilon_{ijk}\sigma^{k} & 0\\
0 & \epsilon_{ijk}\sigma^{k}%
\end{array}
\right)  .
\end{equation}
Note that these components are also expressed as $\sigma^{0j}=i\alpha^{j},$
$\sigma^{ij}=-\epsilon_{ijk}\Sigma^{k}.$ These results are explicitly
evaluated in the following representation of the $\gamma$-matrices:
\begin{align}
\gamma^{0} &  =\left(
\begin{array}
[c]{cc}%
1 & 0\\
0 & -1
\end{array}
\right)  ,\text{ }\gamma^{i}=\left(
\begin{array}
[c]{cc}%
0 & \sigma^{i}\\
-\sigma^{i} & 0
\end{array}
\right)  ,\text{ }\gamma_{5}=\left(
\begin{array}
[c]{cc}%
0 & 1\\
1 & 0
\end{array}
\right)  ,\nonumber\\
& \\
\alpha^{i} &  =\left(
\begin{array}
[c]{cc}%
0 & \sigma^{i}\\
\sigma^{i} & 0
\end{array}
\right)  ,\text{ \ \ \ \ }\Sigma^{k}=\left(
\begin{array}
[c]{cc}%
\sigma^{k} & 0\\
0 & \sigma^{k}%
\end{array}
\right)  .\nonumber
\end{align}
with $\mathbf{\sigma}=(\sigma_{x},\sigma_{y},\sigma_{z})$ being the Pauli
matrices. With this notation, Eq. (\ref{KF}) is written as
\begin{equation}
\left(  K_{F}\right)  _{\mu\nu\alpha\beta}\sigma^{\mu\nu}F^{\alpha\beta
}=2i\alpha^{j}\left(  \mathbb{E}^{j}+\mathbb{B}^{j}\right)  -2\Sigma
^{j}\left(  \mathbb{\tilde{E}}^{j}+\mathbb{\tilde{B}}^{j}\right)  .
\end{equation}
where we have introduced the following definitions:%
\begin{align}
\ \mathbb{E}^{k} &  =\left(  \kappa_{DE}\right)  _{kj}E^{j},\text{
\ \ }\mathbb{B}^{k}=\left(  \kappa_{DB}\right)  _{kj}B^{j},\label{Edef1}\\
& \nonumber\\
\mathbb{\tilde{E}}^{k} &  =\left(  \kappa_{HE}\right)  _{kq}E^{q},\text{
\ }\mathbb{\tilde{B}}^{k}=\left(  \kappa_{HB}\right)  _{kp}B^{p},
\end{align}
and the relation (\ref{Par2}) was used. In the momentum coordinates,
$i\partial_{\mu}\rightarrow p_{\mu},$ the corresponding Dirac equation is
\begin{align}
i\partial_{t}\Psi^{(e)} &  =\left[  \mathbf{\alpha}\cdot(\mathbf{p}%
-e\mathbf{A)}+eA_{0}+m_{e}\gamma^{0}\right.  \nonumber\\
&  \left.  -\lambda^{\left(  e\right)  }i\gamma^{j}\left(  \mathbb{E}%
^{j}+\mathbb{B}^{j}\right)  +\lambda^{\left(  e\right)  }\gamma^{0}\Sigma
^{k}\left(  \mathbb{\tilde{E}}^{k}+\mathbb{\tilde{B}}^{k}\right)  \right]
\Psi^{(e)}.\label{DiracM3}%
\end{align}

\section{Nonrelativistic limit}

In order to investigate the role played by this nonminimal coupling, we should
evaluate the nonrelativistic limit of the Dirac equation. Writing the
spinor\thinspace$\ \Psi$\ in terms of small $\left(  \chi\right)  $\ and large
$\left(  \phi\right)  $\ two-spinors,
\begin{equation}
\Psi=\left(
\begin{array}
[c]{c}%
\phi\\
\chi
\end{array}
\right)  ,
\end{equation}
the Dirac equation (\ref{DiracM3}) leads to two two-component equations,%
\begin{align}
\left[  E-eA_{0}-m_{e}+LV1\right]  \phi-\left[  \boldsymbol{\sigma}%
\cdot(\boldsymbol{p}-e\boldsymbol{A)}-LV2\right]  \chi &  =0,\label{Pauli1}\\
& \nonumber\\[-0.2cm]
\left[  \boldsymbol{\sigma}\cdot(\boldsymbol{p}-e\boldsymbol{A)}+LV2\right]
\phi-\left[  E-eA_{0}+m_{e}-LV1\right]  \chi &  =0. \label{Pauli2}%
\end{align}
with%
\begin{align}
LV1  &  =-\lambda^{\left(  e\right)  }\sigma^{k}\left(  \mathbb{\tilde{E}}%
^{k}+\mathbb{\tilde{B}}^{k}\right)  ,~\\
LV2  &  =i\lambda^{\left(  e\right)  }\sigma^{j}\left(  \mathbb{E}%
^{j}+\mathbb{B}^{j}\right)  .
\end{align}

In this point we notice that the canonical momentum remains defined as
\begin{equation}
\mathbf{\pi}=(\boldsymbol{p}-e\boldsymbol{A)},
\end{equation}
once the term $LV2$ appears with the same sign in Eqs. (\ref{Pauli1},
\ref{Pauli2}). The small component is given by%
\begin{equation}
\chi\simeq\frac{1}{\left[  2m_{e}\right]  }\left[  \boldsymbol{\sigma}%
\cdot\mathbf{\pi}+LV2\right]  \phi,
\end{equation}
which replaced in Eq. (\ref{Pauli1}) leads to%
\begin{align}
&  \left.  \left[  E-eA_{0}-m_{e}+LV1\right]  \phi=\right.  \label{PauliS}\\
&  \hspace{1.7cm}=\frac{1}{\left[  2m_{e}\right]  }\left[  (\boldsymbol{\sigma
}\cdot\mathbf{\pi}\boldsymbol{)}-LV2\right]  \left[  (\boldsymbol{\sigma}%
\cdot\mathbf{\pi}\boldsymbol{)}+LV2\right]  \phi.\nonumber
\end{align}
At first order in the Lorentz violating parameters, the following Hamiltonian
is achieved:%
\begin{align}
H^{\left(  e\right)  } &  =\frac{1}{2m_{e}}\left[  (\boldsymbol{p}%
-e\boldsymbol{A)}^{2}-e\left(  \boldsymbol{\sigma\cdot B}\right)  \right]
+eA_{0}+\lambda^{\left(  e\right)  }\boldsymbol{\sigma\cdot}\left(
\mathbb{\tilde{E}}+\mathbb{\tilde{B}}\right)  \nonumber\\
&  -\frac{\lambda^{\left(  e\right)  }}{m_{e}}\left(  \mathbb{E+B}\right)
\cdot(\boldsymbol{\sigma}\times\mathbf{p})+\frac{e\lambda^{\left(  e\right)
}}{m_{e}}\left(  \mathbb{E+B}\right)  \mathbb{\cdot}(\mathbf{\sigma}%
\times\mathbf{A})\label{HNR}\\
&  +\frac{\lambda^{\left(  e\right)  }}{2m_{e}}\left[  \partial_{i}%
\mathbb{E}^{i}+\partial_{i}\mathbb{B}^{i}+i\boldsymbol{\sigma}\cdot
(\nabla\times\mathbb{E})+i\boldsymbol{\sigma}\cdot(\nabla\times\mathbb{B}%
)\right]  .\nonumber
\end{align}
In the case we deal with uniform fields, the Hamiltonian becomes%
\begin{align}
H^{\left(  e\right)  } &  =\frac{1}{2m_{e}}\left[  (\boldsymbol{p}%
-e\boldsymbol{A)}^{2}-e\left(  \boldsymbol{\sigma\cdot B}\right)  \right]
+eA_{0}+\lambda^{\left(  e\right)  }\boldsymbol{\sigma\cdot}\left(
\mathbb{\tilde{E}}+\mathbb{\tilde{B}}\right)  \nonumber\\
& \nonumber\\
&  -\frac{\lambda^{\left(  e\right)  }}{m_{e}}\left(  \mathbb{E+B}\right)
\cdot(\boldsymbol{\sigma}\times\mathbf{p})+\frac{e\lambda^{\left(  e\right)
}}{m_{e}}\left(  \mathbb{E+B}\right)  \mathbb{\cdot}(\mathbf{\sigma}%
\times\mathbf{A}).\label{HNR3}%
\end{align}

This Hamiltonian induces new effects. Note that the term $\mathbb{E}%
\cdot(\boldsymbol{\sigma}\times\mathbf{p})$ is a generalization of the Rashba
coupling term, $\mathbf{E}\cdot(\boldsymbol{\sigma}\times\mathbf{p}),$ while
$\lambda^{\left(  e\right)  }\left(  \boldsymbol{\sigma\cdot}\mathbb{\tilde
{B}}\right)  $ implies a straightforward tree-level contribution to the
anomalous magnetic moment of the electron. As another example, the term
$\left(  \boldsymbol{\sigma\cdot}\mathbb{\tilde{E}}\right)  $ leads to a kind
of electric Zeeman effect, in the total absence of magnetic field.

\section{Nonrelativistic physical effects}

In this section, we analyze some physical effects induced by the correction
terms enclosed in Eq. (\ref{HNR3}). In this sense, we particularize this
Hamiltonian for some specific configurations of electric and magnetic fields.

We begin discussing the correction induced on the atomic spectrum of Hydrogen.
In order to carry out the contribution associated with the term $\sigma
\cdot\tilde{E}$ involving the spin operator, it is necessary to work with the
wave functions $\Psi_{nljm_{j}m_{s}}^{(e)}=\psi_{nljm_{j}}(r,\theta,\phi
)\chi_{sm_{s}},$ suitable to treat the situations where there occurs addition
of angular momenta ($J=L+$ $S$), with $n,l,j,m_{j}$ being the associated
quantum numbers. In this case, the correction energy is given by:%
\begin{equation}
\Delta_{E}=\lambda^{\left(  e\right)  }\langle nljm_{j}m_{s}%
|\boldsymbol{\sigma\cdot}\mathbb{\tilde{E}}|nljm_{j}m_{s}\rangle.
\end{equation}
Now, we adopt a polarized spin configuration, $\sigma=\sigma_{z}\hat{z},$ such
that%
\begin{equation}
\boldsymbol{\sigma\cdot}\mathbb{\tilde{E}=}\left(  \kappa_{HE}\right)
_{3j}\boldsymbol{\sigma}_{z}E_{j},
\end{equation}
with $E_{j}$ being one of the components of the electric field, and $\left(
\kappa_{HE}\right)  _{3j}$ a non null element of the matrix $\left(
\kappa_{HE}\right)  .$ Thus,%
\begin{equation}
\Delta_{E}=\lambda^{\left(  e\right)  }\left(  \kappa_{HE}\right)  _{3j}%
E_{j}\langle nljm_{j}m_{s}|\boldsymbol{\sigma}_{z}|nljm_{j}m_{s}%
\rangle.\label{EB}%
\end{equation}

To complete this calculation, it is necessary to write the $|jm_{j}\rangle$
kets in terms of the spin eigenstates $|mm_{s}\rangle,$ which is done by means
of the general expression:
\begin{equation}
|jm_{j}\rangle=%
{\displaystyle\sum\limits_{m,m_{s}}}
\langle mm_{s}|jm_{j}\rangle|mm_{s}\rangle,
\end{equation}
where $\langle mm_{s}|jm_{j}\rangle$ are the Clebsch--Gordan coefficients.
Evaluating such coefficients for the case $j=l+1/2,m_{j}=m+1/2,$ one has
$\ |jm_{j}\rangle=\alpha_{1}|m\uparrow\rangle+\alpha_{2}|m+1\downarrow
\rangle;$ one the other hand, for $j=l-1/2,m_{j}=m+1/2,$ one obtains
$|jm_{j}\rangle=\alpha_{2}|m\uparrow\rangle-\alpha_{1}|m+1\downarrow\rangle,$
with $\alpha_{1}=\sqrt{(l+m+1)/(2l+1)},$ $\alpha_{2}=\sqrt{(l-m)/(2l+1)}.$
Taking now into account the orthonormalization relation $\langle m^{\prime
}m_{s}^{\prime}|mm_{s}\rangle=\delta_{m^{\prime}m}\delta_{m_{s}^{\prime}m_{s}%
},$ it is possible to show that Eq.\ (\ref{EB}) leads to:%
\begin{align}
\Delta_{E}  &  =\lambda^{\left(  e\right)  }\left(  \kappa_{HE}\right)
_{3j}E_{j}\langle jm_{j}|\boldsymbol{\sigma}_{z}|jm_{j}\rangle,\\
& \nonumber\\
\Delta_{E}  &  =\pm\lambda^{\left(  e\right)  }\left(  \kappa_{HE}\right)
_{3j}E_{j}\frac{m_{j}}{2l+1}, \label{shift1}%
\end{align}
where the positive and negative signs correspond to $j=l+1/2$ and $j=l-1/2,$
respectively. It was also used $\langle nljm_{j}m_{s}|\sigma_{z}|nljm_{j}%
m_{s}\rangle=m_{j}\hbar/(2l+1)$, $\langle nljm_{j}m_{s}|\sigma_{x}%
|nljm_{j}m_{s}\rangle=\langle nljm_{j}m_{s}|\sigma_{y}|nljm_{j}m_{s}%
\rangle=0.$ The dependence on $m_{j}$ leads to a spectrum splitting in
$\left(  2j+1\right)  $ lines, representing an electric Zeeman-like effect
(due to the presence of an electric field, that can be external or the atomic
one). Regarding the possibility of measuring spectrum shifts as small as
$10^{-10}eV,$ and working with a typical atomic electric field for the
Hydrogen fundamental level ($a_{0}\simeq0.529\mathring{A})$, whose magnitude
is $E\simeq5.1\times10^{11}N/C\simeq1.2\times10^{6}\left(  eV\right)  ^{2},$
the Zeeman-like splitting of Eq. (\ref{shift1}) will be undetectable if
\begin{equation}
\left\vert \lambda^{\left(  e\right)  }\left(  \kappa_{HE}\right)
_{3j}\right\vert E_{j}<10^{-10}\left(  eV\right)  .
\end{equation}
It leads to the following upper bound:%
\begin{equation}
\left\vert \lambda^{\left(  e\right)  }\left(  \kappa_{HE}\right)
_{3j}\right\vert <8\times10^{-17}\left(  eV\right)  ^{-1}. \label{bound1}%
\end{equation}

Now, an observation is worthwhile. In the derivation of the result
(\ref{shift1}) one has used the Hamiltonian (\ref{HNR3}), particularized for
uniform fields. The nucleus Coulombian field, however, is not constant,
opening the possibility of achieving new spectrum shifts stemming from the
varying electric field terms of Eq.(\ref{HNR}), namely, $\nabla\cdot E,$
$\sigma\cdot(\nabla\times E)$ $.$ Knowning the definition (\ref{Edef1}), and
the Coulombian field, $E^{j}=er^{j}/r^{3}$, we obtain
\begin{align}
\nabla\cdot\mathbb{E}  &  =e/r^{3}\left[  \left(  \kappa_{DE}\right)
_{ii}-3\left(  \kappa_{DE}\right)  _{ij}(\cos\theta_{i})(\cos\theta
_{j})\right]  ,\\
\nabla\times\mathbb{E}  &  =\mathbb{-}3e/r^{3}\left[  \epsilon_{ijk}\left(
\kappa_{DE}\right)  _{kp}(\cos\theta_{j})(\cos\theta_{p})\right]  ,
\end{align}
where $\cos\theta_{i}=r^{i}/r.$ We thus note that the expectation value of
\ $\nabla\cdot\mathbb{E}$ only will receive contributions from the trace and
diagonal elements of the symmetric matrix $\left(  \kappa_{DE}\right)  ,$
while the expectation value of $\nabla\times\mathbb{E}$ is affected only by
the nondiagonal terms. In other words, \ the spectrum corrections\ given by
$\int\Psi_{nlm}^{\ast}(\nabla\cdot\mathbb{E})\Psi_{nlm}d^{3}r\ $and $\int%
\Psi_{nlm}^{\ast}(\nabla\times\mathbb{E})\Psi_{nlm}d^{3}r$, are in general
nonnull, with their values being proportional to $\overline{\left(
1/r^{3}\right)  }=\left\langle nlm|1/r^{3}|nlm\right\rangle =[a_{0}^{3}%
n^{3}l(l+1/2)(l+1)]^{-1}.$ Thus, the energy shifts go as $e\left\vert
\lambda^{\left(  e\right)  }\left(  \kappa_{DE}\right)  _{ij}\right\vert
/(ma_{0}^{3})\sim9\times10^{3}\left\vert \lambda^{\left(  e\right)  }\left(
\kappa_{DE}\right)  _{ij}\right\vert (eV)^{2},$ with $e=\sqrt{1/137},$ leading
to the following upper bound, $\left\vert \lambda^{\left(  e\right)  }\left(
\kappa_{DE}\right)  _{ij}\right\vert <1.1\times10^{-14}\left(  eV\right)
^{-1},$ less restrictive than the one of Eq. (\ref{bound1}), however.

Additional Hydrogen spectrum corrections could arise when one considers a
nonminimal coupling for protons similar to the one here devised for electrons.
This can be done proposing (for protons), $D_{\mu}=\partial_{\mu}+ieA_{\mu
}+\frac{\lambda^{(p)}}{2}\left(  K_{F}\right)  _{\mu\nu\alpha\beta}\gamma
^{\nu}F^{\alpha\beta},$ where $\lambda^{(p)}$ concerns the proton
electromagnetic nonminimal interaction. This proposal is analogue to the one
of Refs. \cite{CPT}. Note that $\lambda^{(p)}\neq\lambda^{(e)}$, where
$\lambda^{(e)}$ is related to electron electromagnetic interaction (it is the
constant that appears in Eq. (\ref{covader})). In this case, the modified
Dirac equation is,\textbf{ }%
\begin{equation}
\left[  i\gamma^{\mu}\partial_{\mu}+e\gamma^{\mu}A_{\mu}+\frac{\lambda^{(p)}%
}{2}\left(  K_{F}\right)  _{\mu\nu\alpha\beta}\sigma^{\mu\nu}F^{\alpha\beta
}-M_{p}\right]  \Psi^{(p)}=0,
\end{equation}
where  $\Psi^{\left(  p\right)  }$ is the proton spinor wave function and
$M_{p}$ is the proton mass. We now consider a scenario in which one supposes
simultaneously two nonminimal coupling terms (for the electron and proton
interactions). The\ full fermion Lagrangian is\textbf{ }$\mathrm{{\mathcal{L}%
}}=\mathrm{{\mathcal{L}}}_{(e)}+\mathrm{{\mathcal{L}}}_{(p)},$ where \textbf{
}%
\begin{equation}
\mathrm{{\mathcal{L}}}_{(p)}=\bar{\Psi}^{(p)}(i{\rlap{\hbox{$\mskip 1 mu /$}}}%
\partial+e{\rlap{\hbox{$\mskip 1 mu /$}}}A-M_{p}+\frac{\lambda^{(p)}}%
{2}\left(  K_{F}\right)  _{\mu\nu\alpha\beta}\sigma^{\mu\nu}F^{\alpha\beta
})\Psi^{(p)},
\end{equation}
is the proton Lagrangian and $L_{(e)}$ is given by Eq. (\ref{Le}). Working out
the nonrelativistic limit, one obtains a full Hamiltonian given as
$H=H^{(e)}+H^{(p)},$ where $H^{(e)}$ is the one of Eq. (\ref{HNR}), and
$H^{(p)}$ is the analogue to this one for the proton, with $m_{e}$ replaced by
$M_{p}$ and $-e\rightarrow e.$ It contains new tree-level contributions in
$\lambda^{(p)}$ that yield spectrum corrections. Noticing that $M_{p}%
\simeq1836m_{e},$ the terms of the proton nonrelativistic Hamiltonian
proportional to $M_{p}^{-1}$ will yield bounds less restrictive than the ones
stemming from the electron Hamiltonian at least by a factor $10^{3}$. The
unique term of $H^{(p)}$ able to lead to a competitive bound is $\lambda
^{(p)}\sigma\cdot\left(  \mathbb{\tilde{E}}+\mathbb{\tilde{B}}\right)  ,$
implying the same bound of Eq. (\ref{bound1}), that is, $\left\vert
\lambda^{(p)}\left(  \kappa_{HE}\right)  _{3j}\right\vert <8\times
10^{-17}\left(  eV\right)  ^{-1}.$ A detailed analysis about the spectrum
corrections induced by the Hamiltonian $H=H^{(e)}+H^{(p)}$ seems to be a
sensitive issue for further investigation. 

Another effect enclosed in Hamiltonian (\ref{HNR3}) is concerned with the
anomalous magnetic moment of the electron. A Lorentz-violating study on this
issue was developed in Refs. \cite{Carone}. The electron magnetic moment is
$\boldsymbol{\mu}=-\mu\boldsymbol{\sigma,}$ with $\mu=e/2m_{e}$, and $g=2$ the
gyromagnetic factor. The anomalous magnetic moment of the electron is given by
$g=2(1+a),$ with $a=\alpha/2\pi+...=0.00115965218279$ representing the
deviation \textbf{(}value in the year 2008\textbf{) }in relation to the usual
case. In this case, the magnetic interaction is $\boldsymbol{H}^{\prime}%
=\mu(1+a)\boldsymbol{\sigma}\cdot\mathbf{B}$. In accordance with very precise
measurements and QED calculations \cite{Gabrielse}, precision corrections to
this factor are now evaluated at the level of $1$\ part in $10^{11},$\ that
is, $\Delta a\leq3\times10^{-11}.$\ In our case, the Hamiltonian (\ref{HNR3})
provides tree-level LV\ contributions to the usual $g=2$\ gyromagnetic factor,
which can not be larger than $a$. The total magnetic interaction in Eq.
(\ref{HNR3}) is
\begin{equation}
\frac{e}{2m_{e}}\left(  \boldsymbol{\sigma\cdot B}\right)  +\lambda^{\left(
e\right)  }\left(  \boldsymbol{\sigma\cdot}\mathbb{\tilde{B}}\right)  .
\end{equation}
For the magnetic field along the z-axis, $\mathbf{B=}B_{0}\hat{z},$ and a
spin-polarized configuration in the $z$-axis, this interaction assumes the
form
\begin{equation}
\mu\left[  1+\frac{2m_{e}}{e}\lambda^{\left(  e\right)  }\left(  \kappa
_{HB}\right)  _{33}\right]  \left(  \boldsymbol{\sigma}_{z}B_{0}\right)  ,
\end{equation}
with $\displaystyle\frac{2m_{e}}{e}\lambda^{\left(  e\right)  }\left(
\kappa_{HB}\right)  _{33}$ representing the tree-level LV correction that
should be smaller than $a$. Under such consideration, we obtain the following
upper bound:
\begin{equation}
\left\vert \lambda^{\left(  e\right)  }\left(  \kappa_{HB}\right)
_{33}\right\vert \leq9.7\times10^{-11}\text{ }\left(  eV\right)  ^{-1},
\end{equation}
where we have used $m_{e}=5.11\times10^{5}eV,$ $e=\sqrt{1/137}.$

Finally, we should claim the nonrelativistic Hamiltonian (\ref{HNR3}) possess
a Rashba-like coupling term, $\displaystyle\frac{\lambda^{\left(  e\right)  }%
}{m_{e}}\mathbb{E}\cdot(\boldsymbol{\sigma}\times\mathbf{p}).$ Indeed, the
Rashba spin-orbit interaction (RSOI), given by $H_{R}=\beta_{R}(\sigma
_{x}p_{y}-\sigma_{y}p_{x}),$ has been studied in many works \cite{Rashba}. In
Refs. \cite{RashbaAC}, the usual Rashba coupling term is examined in
connection with quantum transport properties of ring systems, where
Aharonov-Casher effect leads to well defined conductance oscillations. A
recent work has also argued that some terms of the fermion sector of the SME
induces a Rashba-like coupling term \cite{Ajaib}. In accordance with Refs.
\cite{Rashba}, the Rashba constant for a typical mesoscopic system is
$\beta_{R}\simeq10^{-12}(eV\cdot m)=5\times10^{-6}.$ For a typical electric
field ($E\simeq10^{7}$Volt/m$=23.1(eV)^{2}),$ the factor $E/m_{e}$ is
approximately $4\times10^{-5}eV.$ The imposition of the condition
$\frac{\lambda^{\left(  e\right)  }}{m_{e}}\left\vert \mathbb{E}%
\cdot(\boldsymbol{\sigma}\times\mathbf{p})\right\vert <$ $\beta_{R}$ leads to
$\left\vert \lambda^{\left(  e\right)  }\left(  \kappa_{DE}\right)
_{3j}\right\vert <0.1,$ revealing that\ the Rashba coupling phenomenology is
not a good route to constrain this nonminimal coupling.

\section{Conclusions}

We have devised a new CPT-even and Lorentz-violating nonminimal coupling
between fermionic and gauge fields. This dimension-five nonminimal coupling
involves the dimensionless tensor $\left(  K_{F}\right)  _{\mu\nu\alpha\beta}%
$\ which composes the CPT-even and Lorentz-violating electrodynamics of the
SME. It was considered in the context of the Dirac equation and the
nonrelativistic limit was assessed and carried out. The resulting
nonrelativistic Hamiltonian possess new interesting contributions able to
yielding a new effect (a kind of electric Zeeman-like effect), corrections
to\ the anomalous magnetic moment, and a Rashba-like coupling term. The
electric Zeeman-like effect may lead to upper bounds as good as $\left\vert
\lambda^{\left(  e\right)  }\left(  \kappa_{HE}\right)  _{3j}\right\vert
<8\times10^{-17}\left(  eV\right)  ^{-1},$ while the corrections on the
magnetic moment yields an upper bound as tight as $1$ part in $10^{10}.$ It is
important to mention that the bounds here found should not be confused with
the upper bounds on the $\left(  K_{F}\right)  $-CPT-even components already
known in the literature, once in the present case the constraint is on the
magnitude of the CPT-even parameters as nonminimally coupled. Note that
adopting distinct configurations of the electric and magnetic fields (not
along the $z$-axis), similar upper bounds can be imposed on other coefficients
of the matrices $\left(  \kappa_{HE}\right)  $ and $\left(  \kappa
_{HB}\right)  .$

Concerning this dimension-five nonminimal coupling a promising investigation
is related with the radiative corrections stemming from the photon one-loop
vacuum polarization. Carrying out such radiative contributions, we observe
that a dimension-four CPT-even term, $\left(  \lambda^{\left(  e\right)
}m_{e}\right)  (K_{F})_{\mu\nu\rho\sigma}F^{\mu\nu}F^{\rho\sigma},$ is
generated. Dimension-six operators are also generated at second order in
$\lambda^{\left(  e\right)  }K_{F}$. \ Since the dimension-4 operator can be
generated by radiative corrections, the existing bounds \cite{KM1,Risse} on
the CPT-even $\left(  K_{F}\right)  _{\mu\nu\alpha\beta}$ can be used to
achieve even better bounds on the magnitude of the quantity $\lambda^{\left(
e\right)  }\left(  K_{F}\right)  _{\mu\nu\alpha\beta}.$ The fact that this
term has as coefficient $\left(  \lambda^{\left(  e\right)  }m_{e}\right)
$\ allows to attain better bounds on $\lambda^{\left(  e\right)  }K_{F}$\ by
the factor $1/m_{e}\sim10^{-5}$ in comparison with the bounds on $K_{F}.$\ For
instance, typical bounds on the nonbirefringent coefficients, $\left\vert
\lambda^{\left(  e\right)  }K_{F}\right\vert <10^{-18}\left(  eV\right)
^{-1},$\ imply upper bounds as tight as $\left\vert \lambda^{\left(  e\right)
}K_{F}\right\vert <10^{-23}\left(  eV\right)  ^{-1}$\ on the nonminimal
coupling. The detailed analysis of this issue is now under consideration.

Another interesting perspective is concerned with a complete investigation of
the corrections on the Hydrogen spectrum implied by the Hamiltonian
(\ref{HNR}) and $H=H^{(e)}+H^{(p)}$. Such analysis should be carefully carried
out for all the terms involving $E,B,\tilde{E},\tilde{B}$, focusing on the
ones that could yield stringer upper bounds on the LV parameters, and having
as counterpart the procedures of Refs. \cite{NMhydrog}, \cite{Hydrogen}.

Finally, this new coupling may be examined in several distinct respects,
including applications in the ultrarelativistic regime. Very recently, a study
involving an electron-positron scattering in a QED framework endowed with this
nonminimal coupling was successfully performed yielding upper bounds as tight
as\textbf{ } $\lambda^{\left(  e\right)  }\left(  K_{F}\right)  <10^{-12}%
\left(  eV\right)  ^{-1}$\textbf{ }\cite{NMscatt}.

\begin{acknowledgments}
The authors are grateful to CNPq, CAPES and FAPEMA (Brazilian research
agencies) for invaluable financial support.{}
\end{acknowledgments}

\end{document}